\documentclass[fleqn,usenatbib]{mnras}

\usepackage{newtxtext,newtxmath}

\usepackage[T1]{fontenc}

\DeclareRobustCommand{\VAN}[3]{#2}
\let\VANthebibliography\thebibliography
\def\thebibliography{\DeclareRobustCommand{\VAN}[3]{##3}\VANthebibliography}

\usepackage{graphicx}	
\usepackage{amsmath}	
\usepackage{caption}
\usepackage{subcaption}
\def\mnras{{\rm MNRAS}}

\def\hmpc{h^{-1}{\rm Mpc}}

\def\invhmpc{\;h\;{\rm Mpc}^{-1}}

\def\msun{\, M_{\odot}}

\def\simlt{\lower.5ex\hbox{$\; \buildrel < \over \sim \;$}}
\def\simgt{\lower.5ex\hbox{$\; \buildrel > \over \sim \;$}}

\title[SR simulation of FDM Model]{Super-resolution simulation of the Fuzzy Dark Matter cosmological model}

\author[M. Sipp et al.]{
Meris Sipp,$^{1,2}$
Patrick LaChance,$^{1,2}$
Rupert Croft$^{1,2}$
Yueying Ni$^{1,2,3}$ and
Tiziana Di Matteo$^{1,2}$
\\
$^{1}$McWilliams Center for Cosmology, Department of Physics, Carnegie Mellon University, Pittsburgh, PA 15213 USA\\
$^{2}$NSF AI Planning Institute for Physics of the Future, Carnegie Mellon University, Pittsburgh, PA 15213, USA\\
$^{3}$ Harvard-Smithsonian Center for Astrophysics, Harvard University, 60 Garden Street, Cambridge, MA 02138, USA\\
}

\date{Accepted XXX. Received YYY; in original form ZZZ}

\pubyear{2022}

\begin{document}
\label{firstpage}
\pagerange{\pageref{firstpage}--\pageref{lastpage}}
\maketitle

\begin{abstract}
AI super-resolution, combining deep learning and N-body simulations has
been shown to successfully reproduce the large scale structure and 
halo abundances in the Lambda Cold Dark Matter cosmological model.
Here,  we extend its use to models with a different dark matter
content, in this case Fuzzy Dark Matter (FDM), in the approximation that the
difference
is encoded in the initial power spectrum. We focus on
redshift $z=2$, with simulations that model smaller scales and lower masses, the
latter by two orders of magnitude, than has been done in previous AI
super-resolution work. 
We find that the super-resolution
technique can reproduce the power spectrum and halo mass function to within a
few percent of full high resolution calculations.
We also find that halo artifacts, caused by spurious numerical
fragmentation of filaments, are equally present in the super-resolution outputs.
Although we have not trained the super-resolution algorithm using full quantum pressure FDM simulations, the fact that it performs well at the relevant length and mass scales means that it has promise as technique which could avoid the very high computational cost of the
latter, in some contexts.
We conclude that AI super-resolution can become a useful tool
to extend the range of dark matter models covered in mock  catalogs.
\end{abstract}




\section{Introduction}

The most accepted model for structure formation, 
 Cold Dark Matter (CDM) (see e.g., \citealt{arbey21} for a review) has alternatives. Many of these share the
 feature of a lower amplitude than CDM in the power spectrum of
 density fluctuations on galaxy scales (wavenumbers higher than $k\sim10$Mpc$^{-1}$).
This is in order
to address possible small-scale problems with the CDM model
in the event that (relatively poorly understood) hydrodynamics
and star formation processes cannot account for them.
These problems include the overabundance of substructure in CDM galaxies and
cusps at their centers (see comprehensive discussions in e.g., \citealt{weinberg15}
and \citealt{delpopolo17}).
One of the most prominent alternatives to CDM is so called Fuzzy Dark Matter
(FDM, see e.g., \citealt{hu2000}, \citealt{niemeyer20}),
which proposes that dark matter is extremely light, with particle mass on the order of
10$^{-21}$ eV.  Due to this extremely small mass, dark matter has de Broglie
wavelengths that are macroscopic in size, kiloparsec scale rather
than the nanometers of other particles. 
Other models with reduced small scale power include
Warm Dark Matter (WDM, see e.g., \citealt{colombi96}, \cite{irsic17}  ), CDM with a small scale inflationary
cutoff (e.g., \citealt{kamion00}, \citealt{white2000}), and Self Interacting Dark Matter (SIDM, e.g., \citealt{spergel2000}, \citealt{tulin18} ).

In cosmological simulations (see e.g., \citealt{vogel20} for a review)
the physical processes in each these models can be simulated most accurately
using specialized numerical techniques, such as solving the Schrodinger-Poisson
equation (FDM, e.g., \citealt{mocz17}, \citealt{mocz18}, \citealt{may22}), or thermal velocities for particles (WDM, e.g.,
\citealt{pad22}).
An approximation to all these models, which can be accurate enough
for some purposes,  is
merely to modify the initial linear power spectrum, changing it with respect
to CDM on small scales (as in \citealt{bond80}).
For example this approach was used by \cite{ni19} (hereafter N19) for FDM, and by \cite{bode01} and
\cite{smith11} for WDM.
We use  this approximation here, primarily to explore  techniques
for AI Superresolution simulations in a new regime.

As the FDM model has only recently become more prominent
(e.g., \citealt{hui17}), there are fewer high
resolution simulations than for CDM
run with this paradigm \citep[see, for a recent example of deep zoom-in technique] []{Schwabe2022}, which has limited the
studies of small scale structure in the FDM model. 
What is extremely challenging is to couple the large scales, the dynamics of which is dominated by a cut-off in the initial power spectrum (and can be solved using N-body methods) with small scales where the wave interference effects near the de Broglie scale in FDM become important.
For this reason, simulations with statistically meaningful volumes able to
resolve the wave-like dynamics in FDM halos have 
been really limited due to their extreme computational costs.
Through use of AI-assisted
super-resolution techniques these limitations can be
overcome.   AI superresolution (AISR, see e.g., \citealt{werhahn29}, \citealt{jiang20}, \citealt{bode19} ) is a recently
developed computational technique which shows promise in this area, and
may be useful in particular for generating large mock catalogs at
low computational cost.

Super-resolution is a term originally applied to image processing, where
features are added below the original resolution scale (\citealt{lak12}). One technique
for doing this involves the use of Neural Networks (NN, see \citealt{goodfellow16} and \citealt{russell20} 
for reviews) to generate these features,
and this approach was adapted to use
in cosmological simulations by \cite{kodi20} and \cite{li21}.
NN are tool used in  Machine Learning, and in the context of super-resolution,
 an architecture known as a Generative Adversarial Network (GAN, \citealt{goodfellow2014generative})
  has proven useful (\citealt{wang2020}). \cite{li21} and \cite{Ni_2021} have shown that by
 making the output of a GAN conditional on a low resolution cosmological
 simulation, a super-resolution (SR) simulation can be generated.  In those works it was shown  that statistical properties of the simulation, such as the
 power spectrum of mass fluctuations and halo and subhalo mass functions
are within a few percent of their equivalents in a conventionally run
"high resolution" (HR) simulation with the same particle number. The SR models,
not needing to calculate the gravitational dynamics run four orders of
magnitude faster (for a resolution enhancement of a factor of 512 in particle number) than the fully N-body HR simulations.

AISR has so far been used to simulate dark matter
structure formation in the CDM
model on relatively large scales:  the power spectrum P(k)
has been compared to HR simulations up to wavenumber of $k= 16 \invhmpc$ (\citealt{Ni_2021}), and
predictions of AISR simulations for CDM halos and subhalos down to masses of $2.5 \times10^{11} \msun$.  This paper is an incremental step forward - we test AISR simulations
down to smaller
scales and lower masses (two orders of magnitude lower). The aim is to show how we can rapidly run
numerical models with different initial matter power spectra P(k) and use them as tools for inference.

\begin{figure*}
    \centering
    \includegraphics[width=2.0\columnwidth]{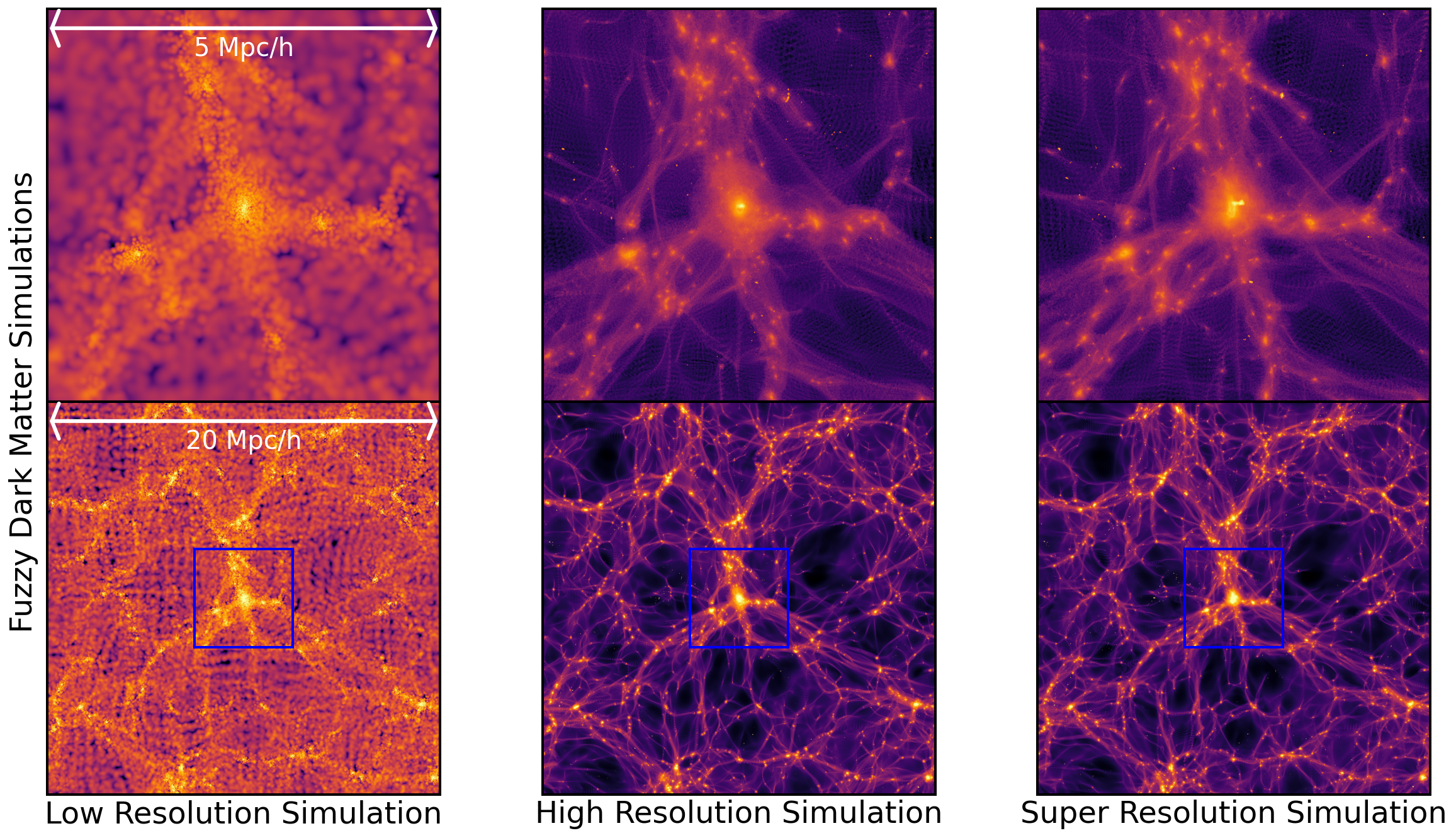}
    \includegraphics[width=2.0\columnwidth]{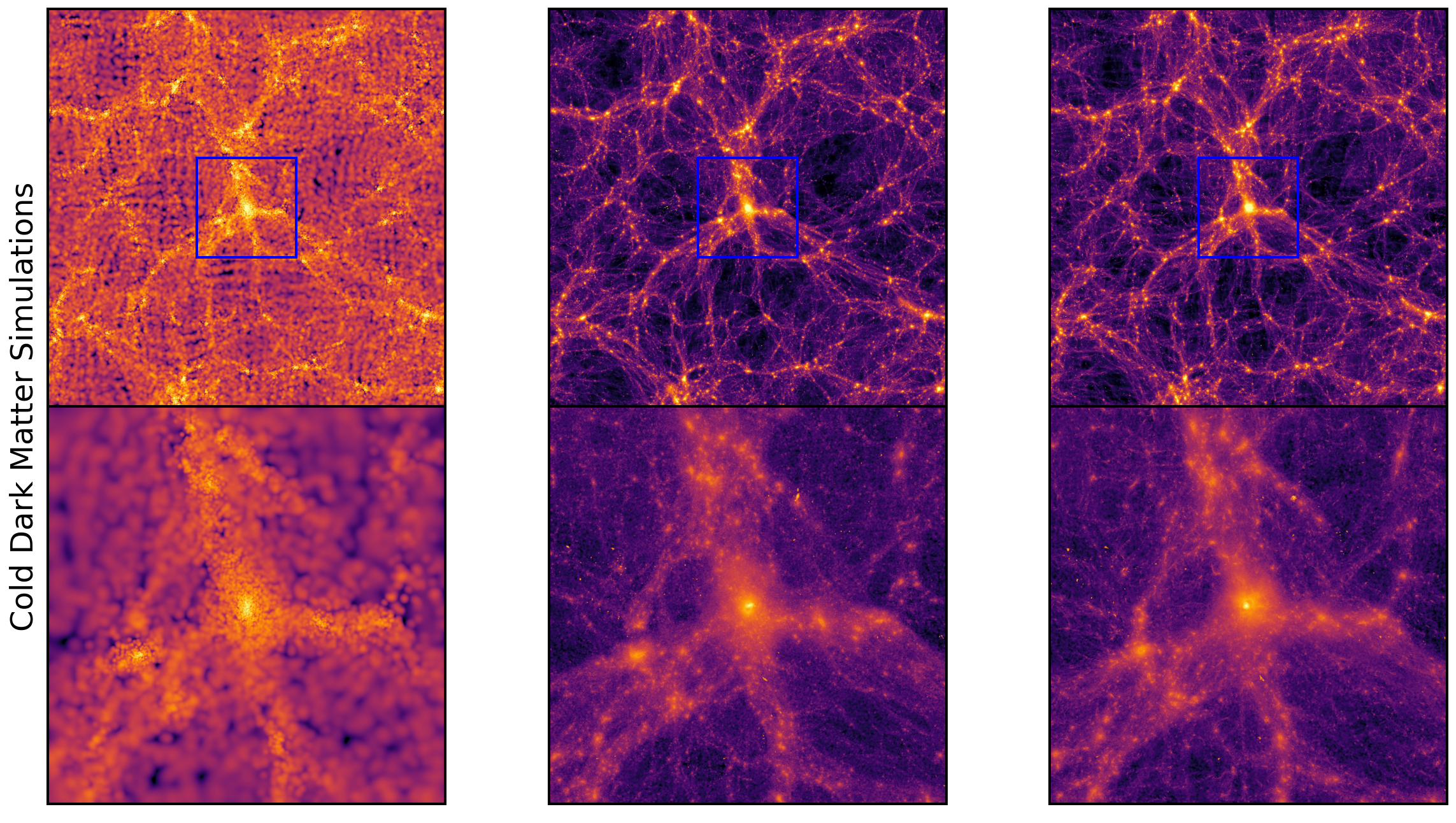}
    \caption{2D visualization of the LR, HR, and SR density distributions for both FDM and LCDM. The central rows show a slice of the enter 20 Mpc simulations, while the top and bottom rows show zoomed in portions of those simulations. The zoomed in locations are indicated with the blue boxes on the central panels. The upper half of the figure is showing the Fuzzy Dark Matter simulations, while the lower half is showing the Cold Dark Matter simulations.}
    \label{fig:Density}
\end{figure*}

\section{Methods}

\subsection{Data}
We work with N-body simulation data for the FDM and LCDM models. Both sets are  run with initial conditions consistent with the WMAP9 cosmology
 with matter density $\Omega _{\rm m} = 0.2814$, dark energy density $\Omega _{\Lambda} = 0.7186$, baryon density $\Omega _{\rm b} = 0.0464$, power spectrum normalization $\sigma_{8} = 0.82$, spectral index $n_{s} = 0.971$, and Hubble parameter $h = 0.697$. The difference between them is 
 solely encoded in the initial power spectrum, with the largest difference being on small scales.
Our FDM simulations employ the initial power spectrum at $z =
99$ generated by {\small AXIONCAMB} (\citealt{hlozek15}), also used in N19.
The FDM particle mass assumed in our models is $2.5\times10^{-22}$ eV, which leads to a damping scale relative to LCDM in the linear P(k)  at
$k=18 \invhmpc$ (see Figure 2 of N19). 

We use the current version of the {\small MP-GADGET} cosmological N-body code (\citealt{bird22}, part of the
{\small GADGET} code family [\citealt{springel05}]) to evolve the initial conditions forward
in time. The simulation is run in dark matter only mode, with particle mass $4.6\times10^{6}\msun$, and a cubical periodic volume of side length $20 \hmpc$. The simulation is evolved from redshift $z=99$ to $z=2$, and we use the final,
$z=2$ snapshot exclusively in our study. We do this
because at high redshifts the differences between the evolved density distributions in FDM and LCDM have not been erased by non-linear evolution (see e.g., \citealt{irsic17}, N19).
Sixteen simulation realizations with
different random seeds are run for each of the FDM and LCDM models. Each of these realizations consists of an LR and and
HR pair. The mass per particle of the LR simulation is
512 times less than for HR, i.e., $2.4\times10^{9}\msun$.

\subsection{Deep Learning Models}
\label{sec:models} 

GANs (\citealt{goodfellow2014generative}) are a deep learning technique that consists of two separate neural networks, named the generative and the discriminative network.
The generative network generates candidate distributions, and the discriminative network tries to determine if that distribution came from an HR simulation (in our case), or from the generator.
In doing this, the GAN learns to generate new data with characteristics matching the training set.
We use a Conditional GAN (CGAN), where the input to the generator
is both noise and a low resolution (LR) version of the simulation data.
The process adds noise during convolution to add irregularity to the smaller scales that was previously absent.
These irregularities become small-scale over and under densities in the distribution of particles of the resulting super-resolution (SR) simulation.

The specific neural network architecture used is that described fully in 
\cite{Ni_2021} and the reader is referred to that paper for more details.
Briefly, the resolution enhancement is applied to the
particle distribution, through the displacement field (displacement of particles from their initial positions). The particle displacements and their
velocities in a low resolution simulation and a noise  vector
are the inputs to the generator. The discriminator network 
receives the high resolution particle positions and displacements, as well as an Eulerian density field representing
those particles assigned to grid cells using a Cloud-in-Cell scheme. The loss function is the total distance between the real and generated datasets using
optimal transport, also known as the Wasserstein distance, 
meaning that the type of GAN we use is a Wasserstein GAN
(WGAN, see \citealt{gul17} for the particular
variant we use). This GAN variation is more stable and
requires less tuning than the original GAN.

\subsection{Training}
\label{sec:methods}

We focus on training our GAN to generate FDM SR particle distributions from LR distributions.
We do this to determine how well a GAN can create a valid SR particle distribution for FDM.
When passing the simulation data to the network for training, we first split it into discrete chunks, due to GPU memory limitations. The chunks have a side length one
fifth of the simulation volume, and the LR simulation input includes an extra margin of the same size, making use of the periodic boundary conditions of the simulation.

After the network is fully trained,
 we generate three different types of SR distributions; one of LCDM from a LR LCDM distribution, one of FDM from a LR FDM distribution and one of FDM from a LR LCDM distribution.
The LCDM to LCDM SR process  follows that employed by \cite{Ni_2021}, except
that it extends to spatial scales 5 times smaller and a mass resolution 125 times
greater. This LCDM SR model is
 used as a baseline for determining the success of the FDM GAN training,
 as well as a test of the SR procedure on smaller scales than have yet been tested. The reader is referred to Section 2.3 of 
 \cite{Ni_2021} for more details.

The FDM to FDM SR distribution was our test case and the LCDM to FDM was used as a proof of concept for seeing if our FDM training could generate an FDM output from an LCDM input.
During training we analysed all of the SR distributions by visual observation and by use of the density field power spectrum, P(k), which is used as a metric to evaluate the SR distribution fidelity. We focus in particular on the region between $k=20\invhmpc$
and $k=150\invhmpc$ as it shows how much small scale structure the SR process has added from the original LR distribution.




\section{Results}

\subsection{Visual impression}

\label{VI}

Fig.~\ref{fig:Density} shows slices of the dark matter density distributions for both LCDM and FDM at all resolutions. We show both a slice through the whole box as well as a smaller region centered on the most massive halo in the volume.
Visually the HR and SR distributions for FDM tend to display less small scale structure as indicated by fewer smaller clumps of particles than the distributions for LCDM as expected from the different models. The LR simulations for both FDM and LCDM are 
very similar in appearance however, and we shall see in Section \ref{secpk} that the most of the extra fluctuation power in LCDM with respect to FDM is below the Nyquist frequency of the
LR model. Comparing the HR and SR distributions for
both models, we have difficulty telling them apart in general appearance, although individual small scale clumps
are different, when looked at closely. As the SR and HR models should only be statistically similar, this is to be expected.

If we look at filaments in the FDM model (for example on the right hand side of the close up panel), we can see that that they are broken up into
regularly spaced clumps, like beads on a string. These small halos are unphysical, as has been noted
in the context of other cosmological models by authors from
\cite{melott89} onwards (including most extensively by \citealt{wang07} and \citealt{angulo13}). These occur most prominently when a model
has little power at frequencies close to the Nyquist frequency of the initial particle grid. If the initial particle distribution is indeed set up in a grid fashion, these fake halos will remain. They are also exacerbated when the numerical resolution is significantly smaller
than the grid spacing (see e.g., \citealt{splinter98}), as is usual in cosmological simulations.
We will see in Section \ref{HMF} when we examine the halo mass function that they are found by the halo finder, and dominate the number of halos below a certain mass. This issue has lead some papers
(e.g., N19) to not use halos below a certain mass in their analyses,
 using the empirically determined halo mass limit for spurious fragmentation
 of \cite{wang07}.  In our case, we note that the SR FDM simulation is sufficiently similar to the HR model that it also includes these spurious objects.

\begin{figure}
     \centering
     \begin{subfigure}[b]{1.0\columnwidth}
         \centering
         \includegraphics[width=\textwidth]{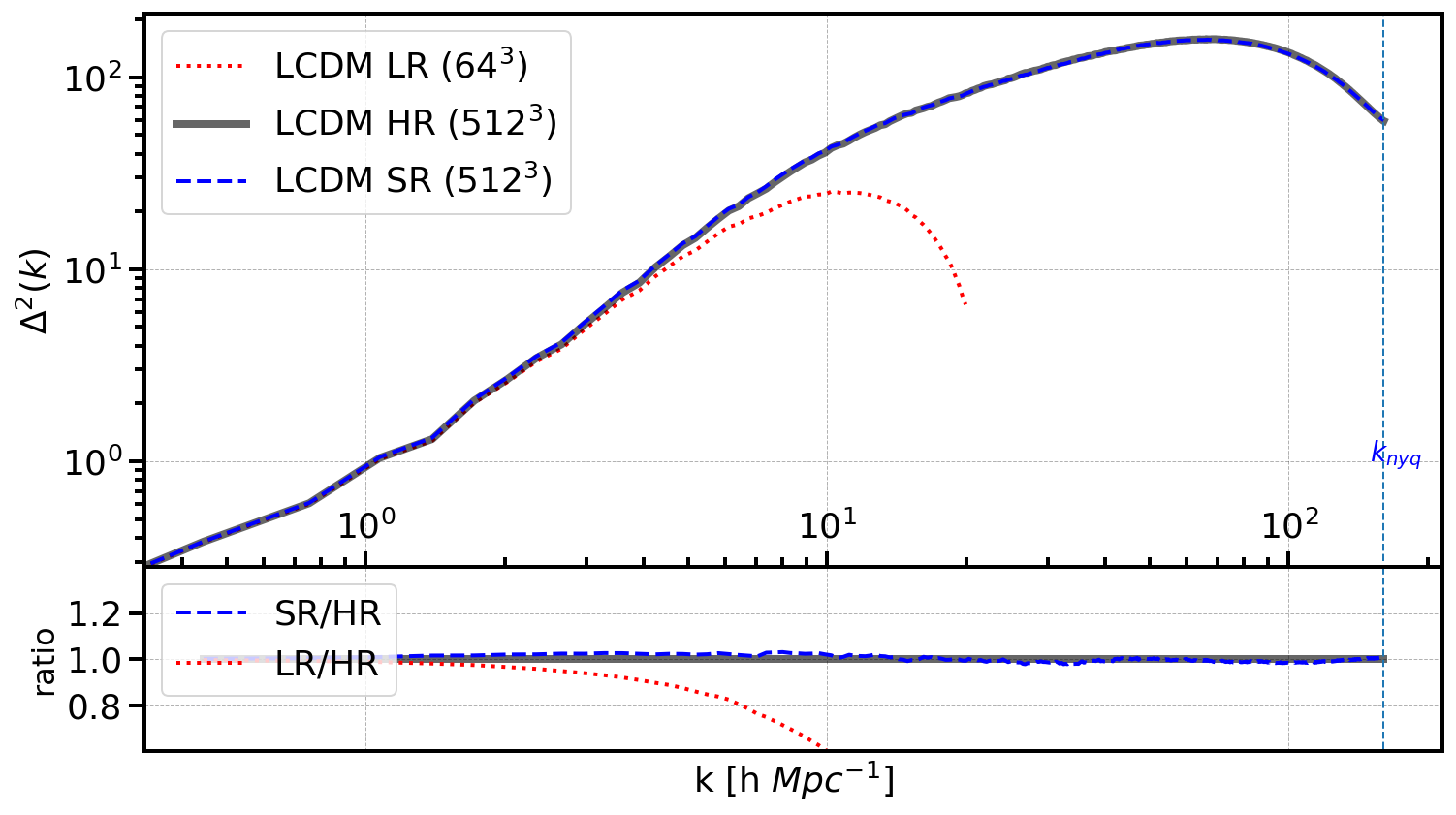}
         \caption{LCDM}
         \label{fig:LCDM_power}
     \end{subfigure}
     \hfill
     \begin{subfigure}[b]{1.0\columnwidth}
         \centering
         \includegraphics[width=\textwidth]{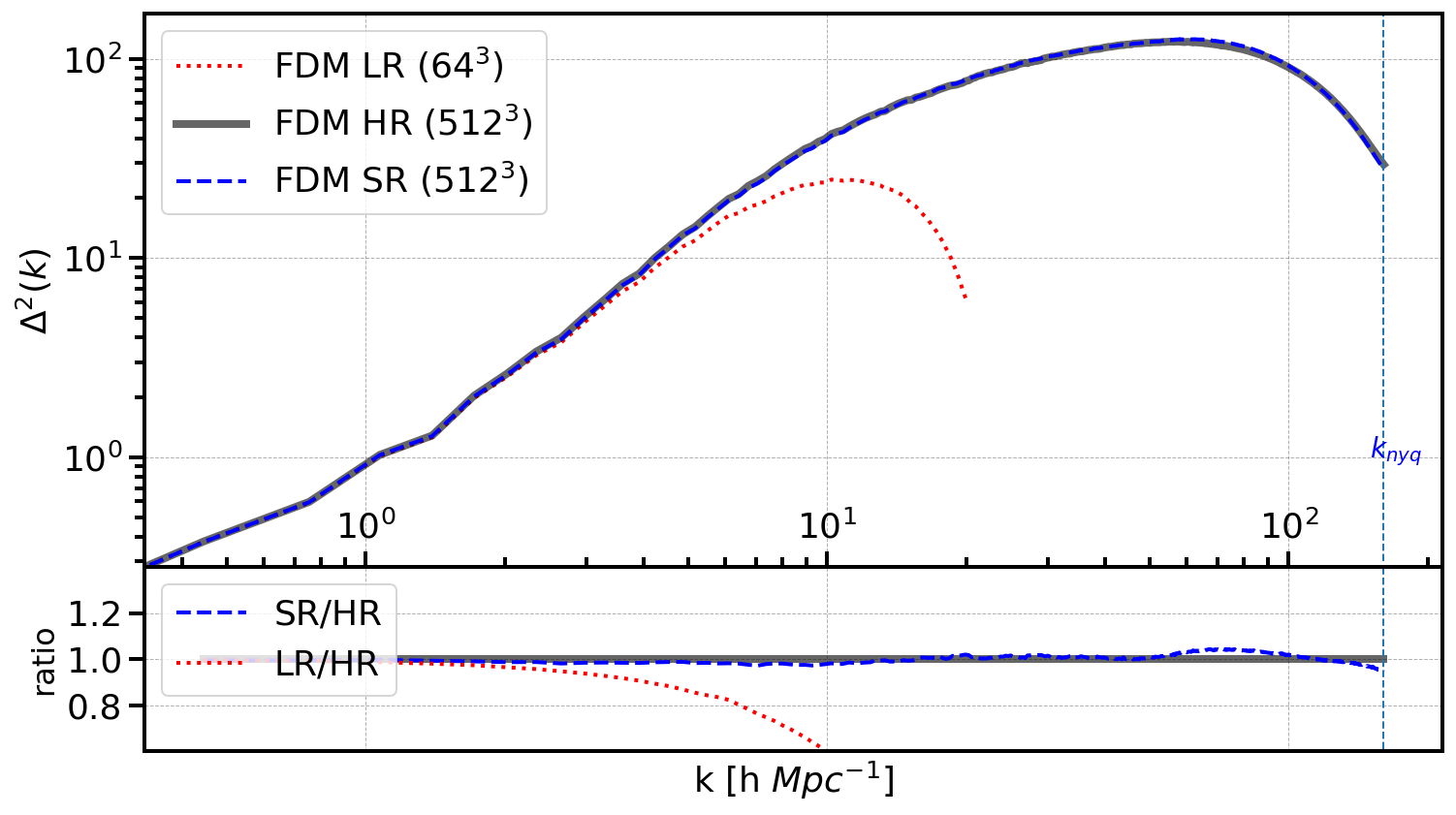}
         \caption{FDM}
         \label{fig:FDM_power}
     \end{subfigure}
     \hfill
     \begin{subfigure}[b]{1.0\columnwidth}
         \centering
         \includegraphics[width=\textwidth]{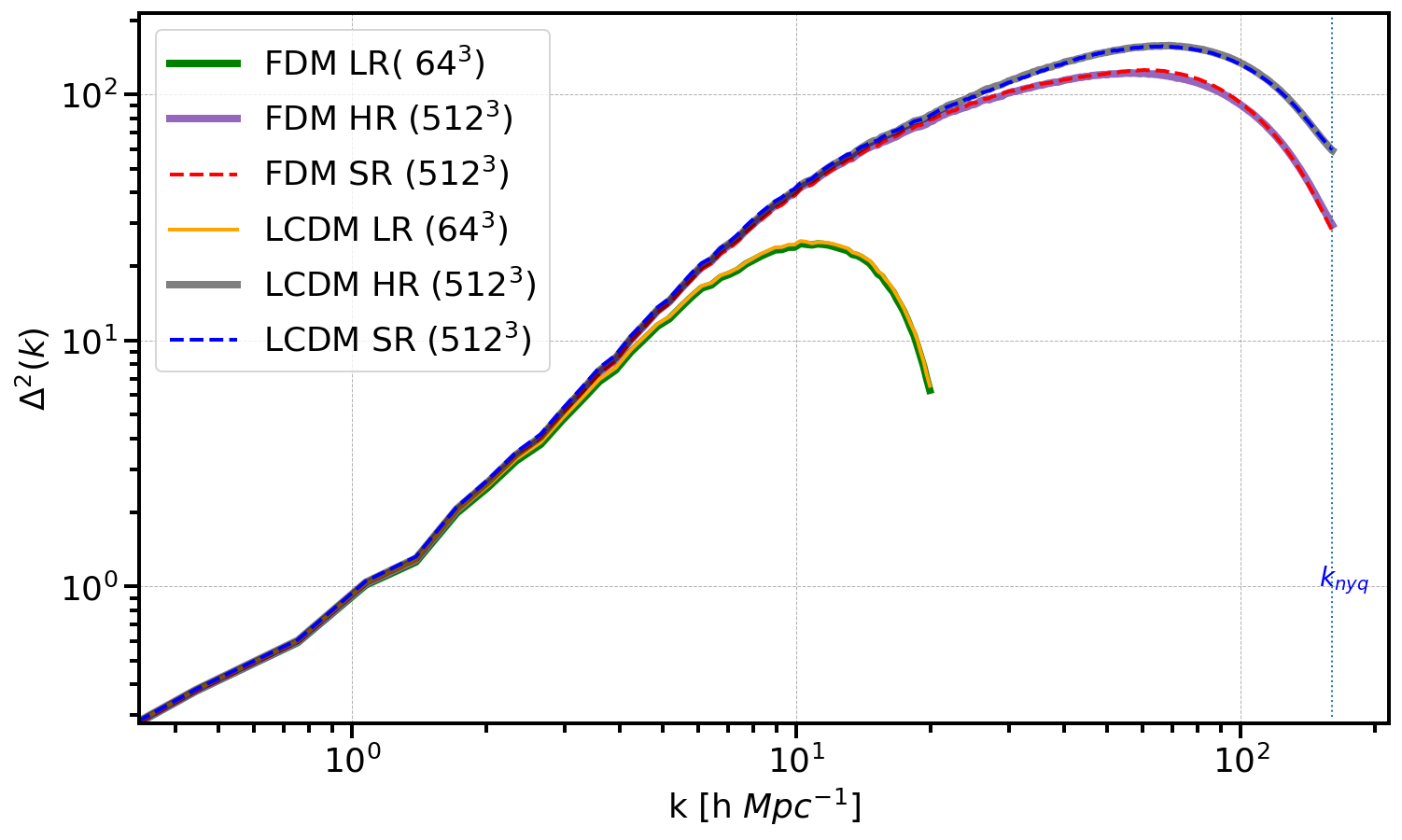}
         \caption{Combined}
         \label{fig:all_power}
     \end{subfigure}
        \caption{The power spectra for both LCDM and FDM. (a) gives the LCDM dimensionless matter power spectrum $\Delta^2$ computed for LR (dotted red), HR (black), and SR (dashed blue). The bottom panel gives the ratio of of both LR (dotted red) and SR (dashed blue) to HR (b) displays the same for FDM. (c) displays all power spectra for both LCDM and FDM on a combined plot. The dashed black line on the right of the figure marks the Nyquist frequency.
        }
        \label{fig:power}
\end{figure}

\subsection{Power spectrum}

\label{secpk}
We compute the power spectra, $P(k)$ for all simulations after assigning the particles
to a grid with the same dimensions as the intial particle grid.
In Fig.~\ref{fig:power}, the dimensionless  power spectra ($\delta^{2}(k)=k^{3}P(k)$ 
for both dark matter models are shown,
with all results being at our chosen redshift of $z=2$.
We plot all curves up to the Nyquist frequency of the grid, including for the LR simulations, for which this is a factor of 8 lower than for HR and SR.
The LCDM model at this
redshift has more power on small scales than FDM, 50\% less at the Nyquist frequency,
the smallest scale plotted.
The SR power spectrum matches the HR spectrum within 5\% on all scales for both FDM and LCDM indicating the GAN was successfully trained. 

\begin{figure}
     \centering
     \begin{subfigure}[b]{1.0\columnwidth}
         \centering
         \includegraphics[width=\textwidth]{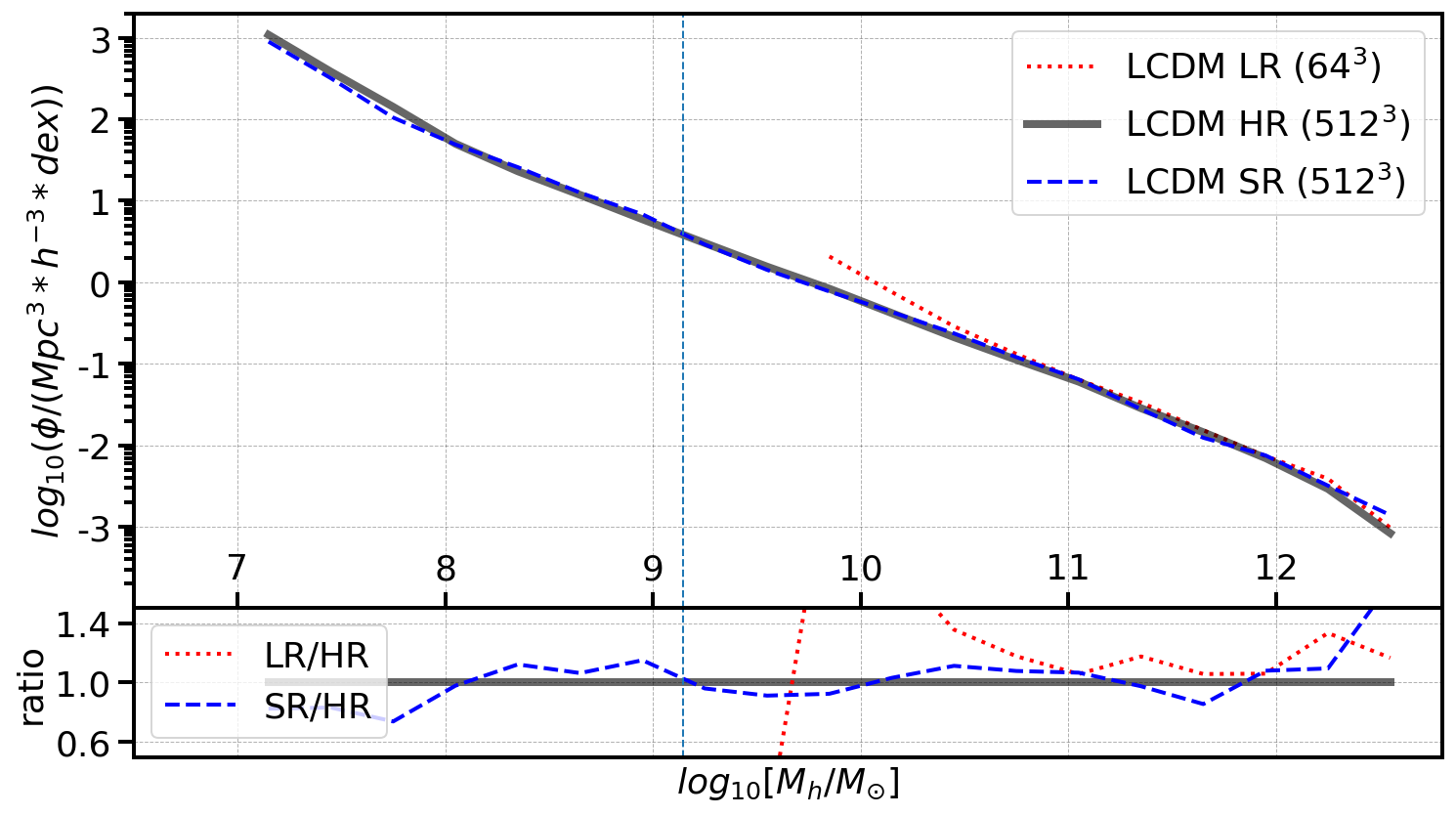}
         \caption{LCDM}
         \label{fig:LCDM_HMF}
     \end{subfigure}
     \hfill
     \begin{subfigure}[b]{1.0\columnwidth}
         \centering
         \includegraphics[width=\textwidth]{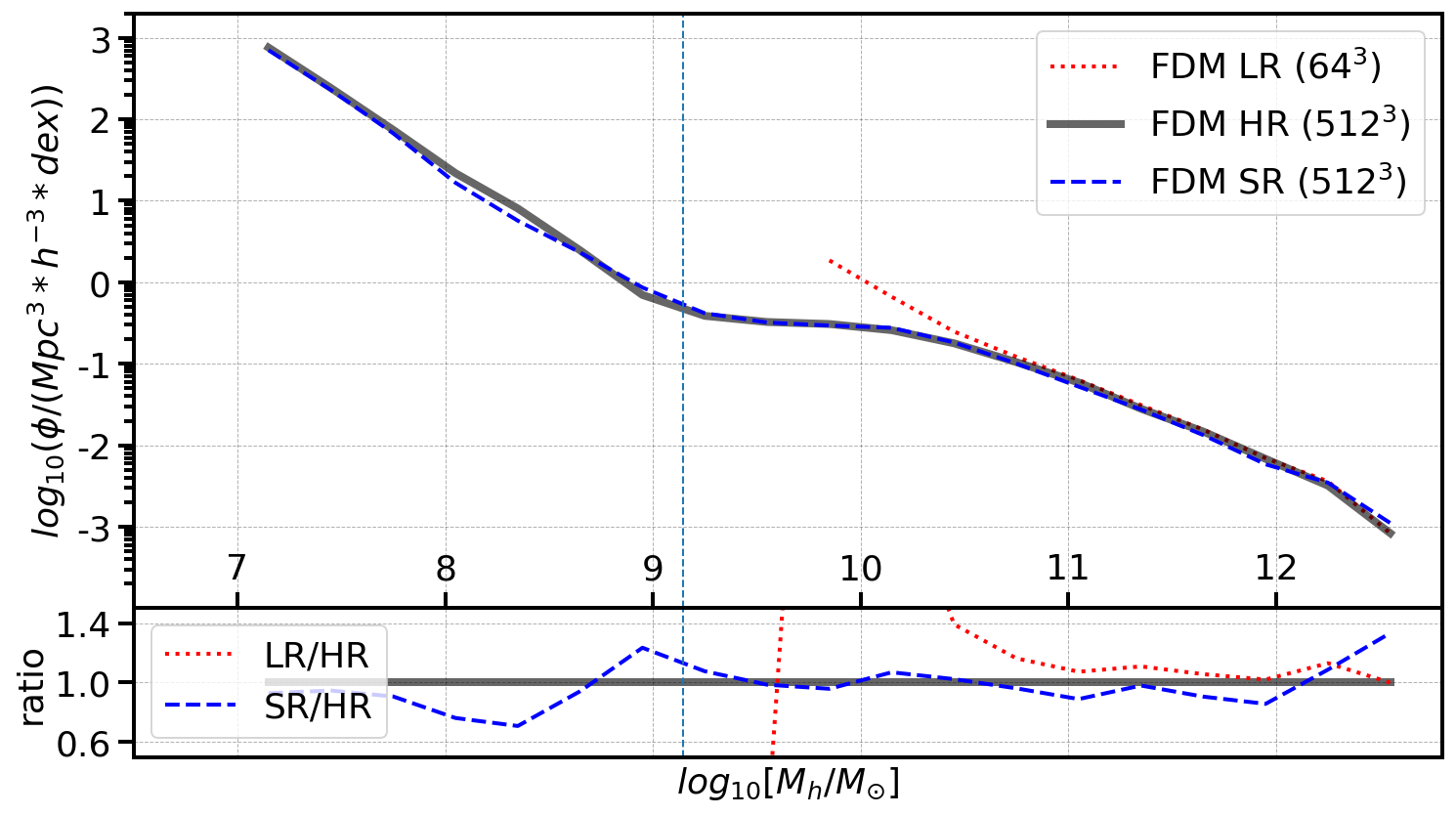}
         \caption{FDM}
         \label{fig:FDM_HMF}
     \end{subfigure}
     \hfill
     \begin{subfigure}[b]{1.0\columnwidth}
         \centering
         \includegraphics[width=\textwidth]{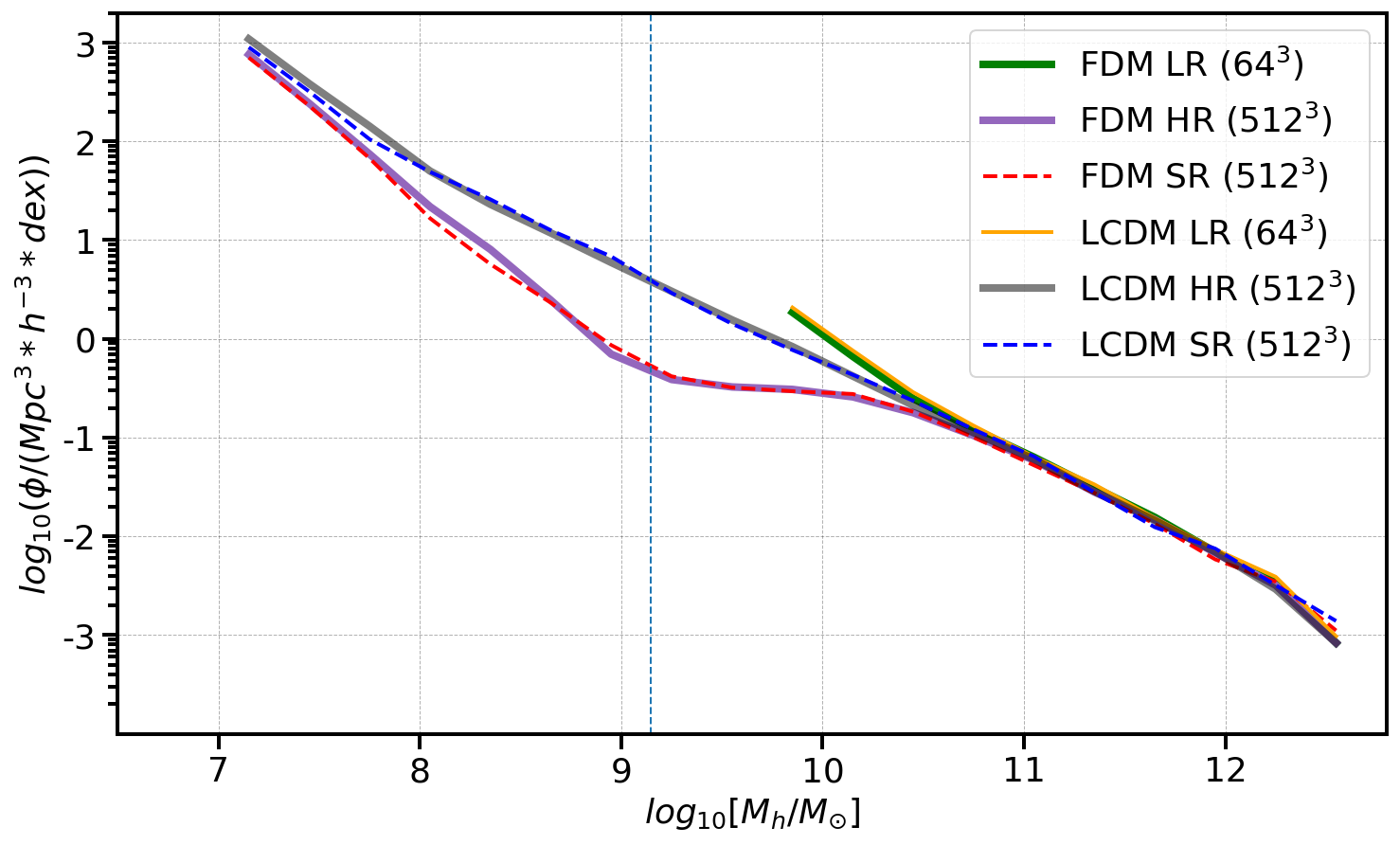}
         \caption{Combined}
         \label{fig:all_HMF}
     \end{subfigure}
        \caption{The halo mass functions for LCDM and FDM. (a) gives the LCDM halo mass functions of the LR (dotted red), HR (black) and SR (dashed blue). The bottom panel gives the ratios of LR (dotted red) and SR (dashed blue) to HR. (b) displays the same for FDM. (c) displays all mass functions for both LCDM and FDM on a combined plot. The vertical dashed grey line indicates the 300 particle cutoff for halo size.}
        \label{fig:HMF}
\end{figure}

\subsection{Halo mass function}
\label{HMF}

We make halo catalogues from the simulation output, using a Friend-of-Friends
halo finder (\citealt{davis85}), with a linking length of 0.2 times the mean
interparticle separation. Following N19 we choose 300 particles as the cutoff below which halos are not simulated
reliably. Numerical artefacts strongly affect the small halo mass function in models 
with low power on halo scales as mentioned in Section \ref{VI} above.

In Figure~\ref{fig:HMF}, we show the halo mass functions are given for both LCDM and FDM.  We can see that the SR and HR mass functions match within 5\% for both FDM and LCDM. Near the 300 particle cutoff, which corresponds to
halo masses of $1.4\times 10^{9} \msun$, we can see that the HR and SR mass functions for FDM are less than those of LCDM by an order of magnitude. This corresponds to a deficit
of low mass galaxies with respect to LCDM which could be 
used to constrain the FDM particle mass (e.g., N19). At redshift $z=2$ which we are plotting here, the deficit is smaller than at higher redshifts, but
still substantial for the relatively low FDM mass we
are simulating ($2.5 \times 10^{-22}$ eV).

We plot halos down to 2 particles, a mass of $9\times10^{6}\msun$ (for HR and SR). Even though most of the halos below 300 particles are not
reliably simulated halos, due to both numerical problems with
FDM/WDM (as mentioned above) and shot noise, we can see
that both the SR and HR simulation results track each other closely all the way down to the lowest masses plotted.  Because the SR simulation process works with the particle displacements and velocities it effectively
generates simulation data in all its details, including 
the spurious halos that were present in the training set.

\begin{figure}
     \centering
     \begin{subfigure}[b]{1.0\columnwidth}
         \centering
         \includegraphics[width=\textwidth]{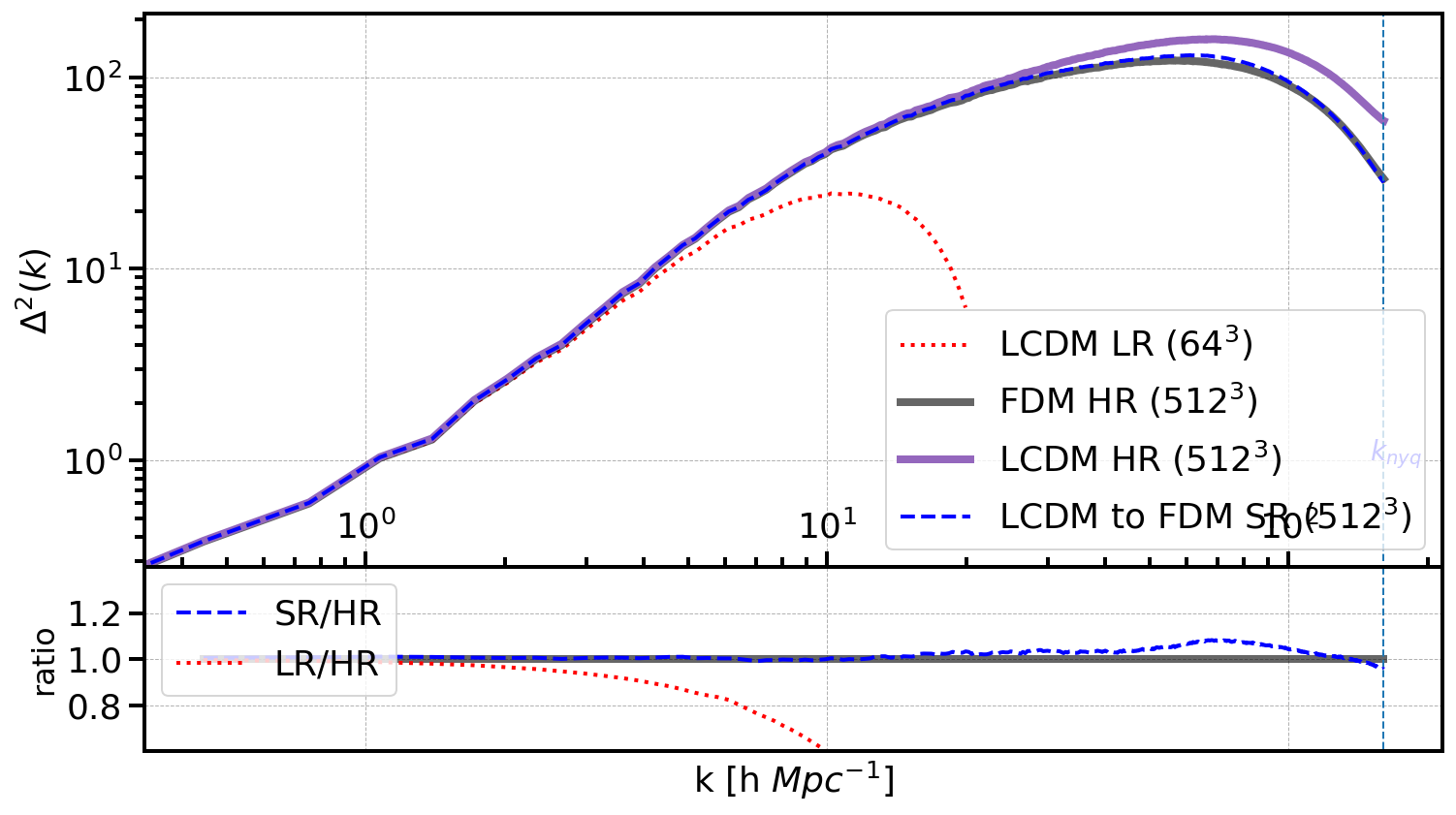}
         \caption{Power Spectrum}
         \label{fig:LCDM_to_FDM_PS}
     \end{subfigure}
     \hfill
     \begin{subfigure}[b]{1.0\columnwidth}
         \centering
         \includegraphics[width=\textwidth]{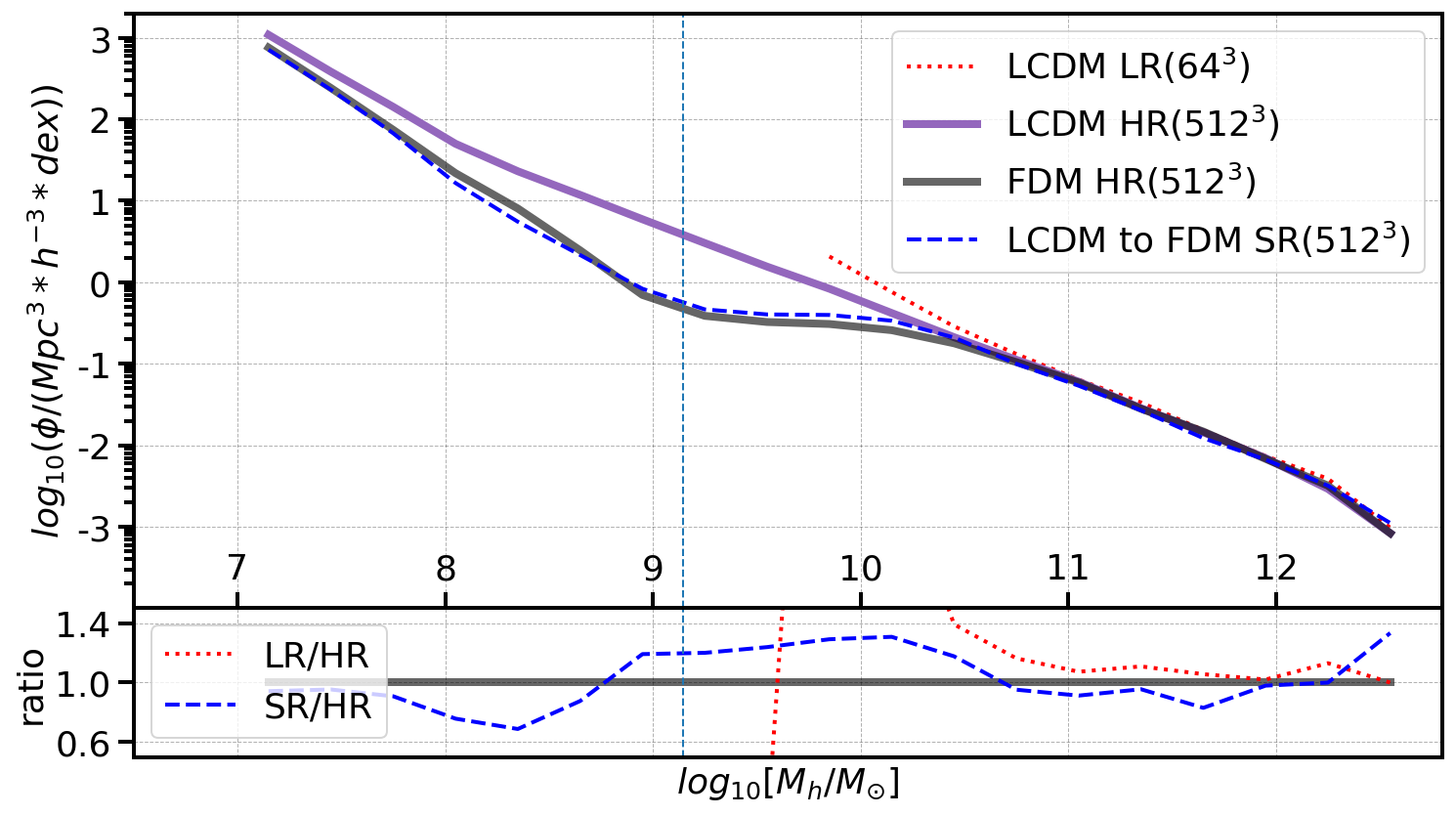}
         \caption{Halo Mass Function}
         \label{fig:LCDM_to_FDM_HMF}
     \end{subfigure}
     
        \caption{Summary statistics for an FDM SR simulation using LR from a different model (LCDM) as an input (see Section
        \ref{hrd}). (a) P(k) for the different models, as in Figure \ref{fig:power}, with the FDM HR using an LCDM LR input shown as a dashed line. (b) The halo mass functions for the
        different models, as in Figure \ref{fig:HMF}.}
        \label{fig:LCDM_to_FDM}
\end{figure}

\subsection{SR of FDM using LCDM LR}
\label{hrd}

The LCDM and FDM simulations have very similiar initial (linear) power spectra
at scales larger than the damping scale ($k=18 \invhmpc$ in the case of $m_{\rm FDM}=2.5\times10^{-22}$ eV, see Figure 2 of N19.). This scale
is above the Nyquist frequency of the LR simulations,
and so opens the possibility of using LR simulations to make SR simulations for a different model.
In our final test we have tried using the LR LCDM simulations as an input to the FDM-trained network.
The results are shown in Figure \ref{fig:LCDM_to_FDM_PS}, where we can see that the resulting
SR FDM power spectrum is quite close to that of the HR FDM model, with a maximum deviation 
of 8\%. The halo mass function, in  Figure \ref{fig:LCDM_to_FDM_HMF} is also 
within 20\% of the FDM HR results.

These tests show that we could in principle use the same LR simulations to 
generate different SR dark matter models, as long as the differences
in the power spectra are on small enough scales and the accuracy is sufficient for the desired use case (e.g., mock catalogues). The NNs in the present case
are trained on a specific DM model, however, and so require full HR simulations
of that a model. A more general tool would be able to train on a range
of models and interpolate between them. The NN architecture StyleGAN2
(e.g., \citealt{karras19}) has shown some promise for field level emulation
in different cosmological models (see e.g., \citealt{jamieson22}) and we hope to 
explore its use in super-resolution in future work.

\section{Summary and Discussion}

\subsection{Summary}
This work attempts to extend the use case of the super resolution architecture developed in \cite{li21} beyond the LCDM model.
The results indicate this extension works well for the case of FDM.
Other viable alternative DM models, such as self-interactive DM \citep{Vogelsberger2012} or warm dark matter \citep{bode01} also feature suppression of the linear matter power spectrum.
Extensions like this indicate that the approach is ripe for further development to allow a single trained network to be able to produce simulations with different cosmological models, which would be an extremely useful tool in studying possible alternate cosmological models.

\subsection{Discussion}
This paper represents a small step on the way to the exploitation
of AISR techniques in cosmology. We have shown that dwarf galaxies and
subhalos with masses of $10^{9} \msun$ can be quite accurately simulated, even for
the LCDM model, which has significant small scale structure, with
orbit crossing and virialization obviously being very important on these
scales. The models still only include dark matter, and the real test will
come when methods to include hydrodynamics are developed. Hydrodynamics will
also break the approximate scale invariance of the LCDM model on
these scales and provide an even more strenuous test of the
AISR technique.

An application of FDM simulations is to compare to the
abundance and statistical properties of galaxies and therefore constrain
the FDM particle mass. At present some of the tightest constraints on this parameter
come from measurements of the Lyman-alpha forest power spectrum (e.g., \cite{irsic17_2}.
N19 have shown that the high redshift galaxy population can also provide
competitive constraints, and with different possible systematic errors
(for example the Lyman-$\alpha$ forest is also sensitive to the
temperature of the intergalactic medium). In this paper we have used
simulations at the relatively high redshift of $z=2$ because non-linear
evolution to lower redshifts tends to erase the differences between
CDM and FDM. At higher redshifts relevant for JWST observations
(e.g., $z=7-12$, \citealt{marshall22}, \citealt{bradley22}), N19 have shown using hydrodynamic simulations that there are significant differences
in the galaxy stellar mass functions between an FDM and LCDM models.
In our simulations, the maximum $\sim$ 5\% differences between HR and SR are smaller
than those between LCDM and FDM models with mass of $2.5\times10^{-22}$ eV. As shown by
\cite{li21}, the accuracy of SR increases at higher redshift, meaning
that heavier mass FDM models will be feasible to simulate with FDM.

Ideal applications for the AISR techniques will be studies of observational
systematics using large mocks, where the uncertainties modelled are
larger than the demonstrated accuracy of AISR. In a similar vein,
semi-analytic models could be run on AISR FDM simulations, where the
uncertainties in the model parameters are larger than AISR accuracy. The
speed of AISR would allow many tens of thousands mock catalogs to be
generated for the same computational cost of a single HR model. We have also
shown that for models such as CDM where the large-scale P(k) is close to
that of FDM, it is even possible to use the same LR simulation to
make SR models with different particle masses, thereby saving on the
cost of running the larger LR N-body simulation. 

Because we have used a very simplified treatment of FDM, the truncation
of the input power spectrum, our conclusions about the success of AISR
also apply to other models when this approximation to the small scale
behaviour is used, such as WDM. There is however nothing in principle  to
prevent the use of AI SR techniques together with more physically accurate 
methods for simulating structure formation in these models.  The methods
used by e.g., \cite{mocz17} and \cite{may21} to solve the Schrodinger-Poisson
equation could be used to train the NN used in AISR. 
This could be used to produce rapid simulations that reproduce the
quantum pressure effects seen by e.g., \cite{may22} beyond those linked
directly to the suppression of power.
We note that
particle discreteness
problems that cause fake substructures to form in models with low small
scale power are also present in our current AIRS model results
(and are visible in Figures \ref{fig:Density}). These could be corrected with
better training simulations and analysis  (e.g., \citealt{angulo13}, \citealt{may21}).

\section*{Acknowledgements}
MS acknowledges support from the NSF AI Institute: Physics of the Future Summer Undergraduate Research Program in Artificial Intelligence and Physics.
RACC and TDM  are supported by  NASA ATP 80NSSC18K101,  NASA ATP NNX17AK56G, and the NSF AI Institute: Physics of the Future, NSF PHY- 2020295. RACC also acknowledges support from NSF AST-1909193 and TDM from
 NSF ACI-1614853 and NSF AST-1616168.

\section*{Data Availability}

 The 
SR models and the pipeline used to generate the SR fields are
available at https://github.com/yueyingn/SRS-map2map.
 The simulation outputs
 and measured summary
 statistics are available
 on request from the authors.



\bibliographystyle{mnras}
\bibliography{refs}




\bsp	
\label{lastpage}
\end{document}